\documentclass[twocolumn,secnumarabic,amssymb, nobibnotes, aps, prl]{revtex4-1}

\usepackage{graphicx}

\divide\pdfpxdimen by 600

\begin{document}
\title{Toward phonon-boundary engineering in nanoporous materials}%

\author{Giuseppe Romano}%
\email[Email: ]{romanog@mit.edu} 
\author{Jeffrey C. Grossman}%
\email[Email: ]{jcg@mit.edu} 
\affiliation{Department of Materials Science and Engineering, Massachusetts Institute of Technology,  77 Massachusetts Avenue, Cambridge (MA), 02139}
\date{Jan 7, 2014}
\begin{abstract}

Tuning thermal transport in nanostructured materials is a powerful approach to develop high-efficiency thermoelectric materials. Using a recently developed approach based on the phonon mean free path dependent Boltzmann transport equation, we compute the effective thermal conductivity of nanoporous materials with pores of various shapes and arrangements. We assess the importance of pore-pore distance in suppressing thermal transport, and identify the pore arrangement that minimizes the thermal conductivity, composed of a periodic arrangement of two misaligned rows of triangular pores. Such a configuration yields a reduction in the thermal conductivity of more than $60 \%$ with respect the simple circular aligned case with the same porosity. 

\end{abstract}
\maketitle

Engineering thermal transport in nanostructured materials is a powerful route to develop high-efficiency thermoelectric (TE) materials. In fact, thanks to phonon-boundary scattering, the effective phonon thermal conductivity (PTC) may decrease even by two orders of magnitude with respect to bulk~\cite{majumdar2004thermoelectricity,dresselhaus2007new,lee2008nanoporous,hochbaum2008enhanced}. Among different nanostructures appealing for TE, porous materials offer the highest number of degrees of freedom for tuning thermal transport, thanks to the possibility of arranging arbitrary boundaries against which phonons can scatter~\cite{lee2008nanoporous,hopkins2009origin,song2004thermal,romano2012mesoscale,hsieh2012thermal}.  In exploring different pore configurations, Song \textit{et al.}~\cite{song2004thermal} found experimentally that for cylindrical pores, the staggered configurations have slightly lower PTCs than the aligned configurations. In other work,   Tse-yang \textit{et al.} provide insights on the influence of the shape of aligned pores on the PTC, by employing the frequency-dependent Boltzmann transport equation (FD-BTE)~\cite{hsieh2012thermal}. In investigating thermal transport across periodically aligned circular pores, Prasher emphasized the importance of the the so-called \textit{view factor}, \textit{i.e.} the proportion of phonons flux leaving the hot surface that strikes the cold surface ballistically~\cite{prasher2006transverse}. However, despite these important contributions, both theoretically and experimentally, in understanding heat transport in complex-shaped porous materials, the combined effects of pores shape, size and arrangement on the PTC is still to a large extend poorly understood. 

In this work, we compute phonon size effects in nanoporous materials with pores of different shapes including circles, squares and triangles, arranged in both aligned and staggered configurations. From our calculations, we deduce that the pore-pore distance plays a crucial role in tuning the PTC, while the view factor plays a secondary role. By using this finding, we identify the optimum pore configuration composed of two misaligned rows of triangular pores, which suppresses heat more effectively than the simple aligned pore configuration case by a factor of $60\%$. Our findings provide practical guidance for engineering thermal transport in nanostructured materials.

Although our approach could be useful for a range of thermal transport related applications, we focus here on TE materials. The TE figure of merit is given by $ZT=\frac{S\sigma^2 T}{\kappa_{eff}}$, where $\sigma$ is the electronic conductivity, $S$ the Seebeck coefficient, $\kappa_{eff}$ the PTC and $T$ the temperature. For simplicity, we neglect the thermal conductivity arising from electrons and assume that the Seebeck coefficient is the same as in bulk. For heavily doped semiconductors, the electron mean free path (MFP) is smaller than $10 nm$~\cite{ziman2001electrons},  thus the drift-diffusion theory for electron transport can be safely used. Within this assumption, the reduction in conductivity of electrons is comparable with that experienced by the heat carriers at the diffusive level, and both reductions can be described by the so called \textit{porosity function}~\cite{hashin1962variational}. As we will see later, for a given porosity, the actual pore configuration has little effect on the porosity function. On the other hand,  phonons in certain materials, such as Si, may have MFP even longer than $1 \mu m$~\cite{esfarjani}; hence, by choosing accurately the characteristic length of the material, it is possible to suppress phonon transport while leaving the electrical transport essentially unaltered and consequentially increasing $ZT$~\cite{dresselhaus2007new}. In light of the discussion above and with minimal loss of generality, in this work we focus on nanoporous (np)-Si with a fixed porosity $\phi = 0.25$.

Quasi-ballistic heat transport is captured by solving the phonon mean free path (MFP)-BTE~\cite{romano_prl} over a simulation domain composed by a square single cell of length $L=10nm$ containing one pore in the center.  Staggered configurations have a square single unit cell of size $L_s=L \sqrt{2}$ with a pore in the center and four pores at the edges.  Hereafter we refer to aligned circles, squares and triangles cases as CA, SA and TA, respectively, whereas the staggered configurations are referred as CS, SS and TS, respectively. We consider the reduction in thermal transport only given by phonon-boundary scattering, whereas the coherence effects are neglected~\cite{PhysRevB.89.205432}. As a consequence, the full phonon dispersion as well as the scattering times are assumed to be the same as in bulk. The PTC is computed by $\kappa_{eff}  = \int_0^\infty K^{bulk}(\Lambda) S(\Lambda) d\Lambda$, where $K^{bulk}(\Lambda)$ is the bulk MFP distribution, obtained by-first principle calculations~\cite{esfarjani}, and $S(\Lambda)$ is the so-called phonon suppression function, the output of the MFP-BTE~\cite{romano_prl}. 

\begin{figure}[htb]
\begin{center}
\includegraphics[width=7cm]{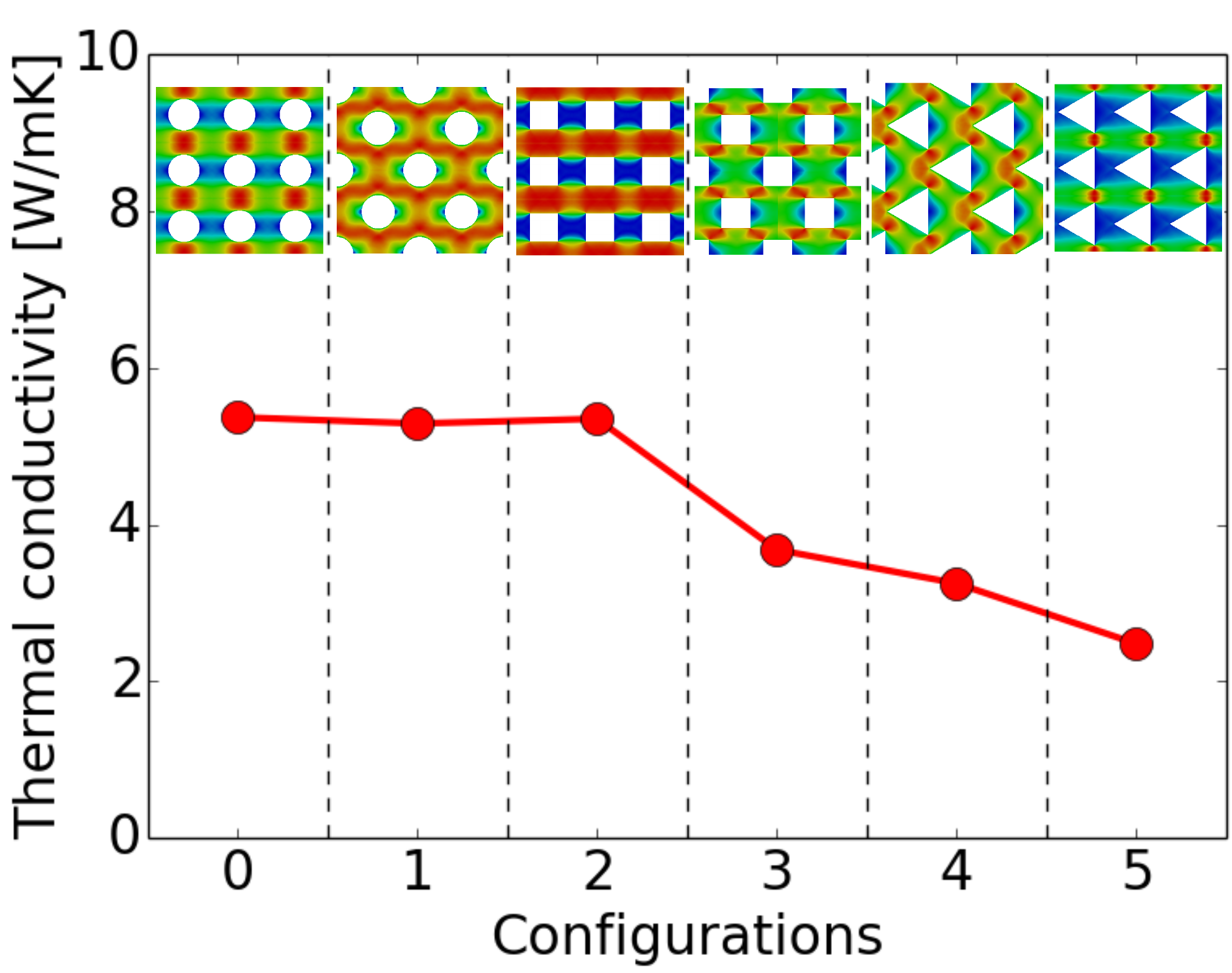}
\caption{Phonon thermal conductivities (PTC) for different configurations. The CA configuration gives the highest PTC whereas the lowest PTC is given by the TA case.   }\label{Fig:kappa}
\end{center}
\end{figure}

To ensure phonon flux, we impose the temperatures $T_{hot}$ and $T_{cold}$ to the left and right side, respectively, of the simulation domain. We apply periodic boundary conditions both along the direction perpendicular and longitudinal to the heat flux. The surface of the pores are considered purely diffusive. The technical details of the implementation of the BTE are in~\cite{romano2011multiscale,romano2012mesoscale}. 
\begin{figure}[htb]
\begin{center}
\includegraphics[width=7cm]{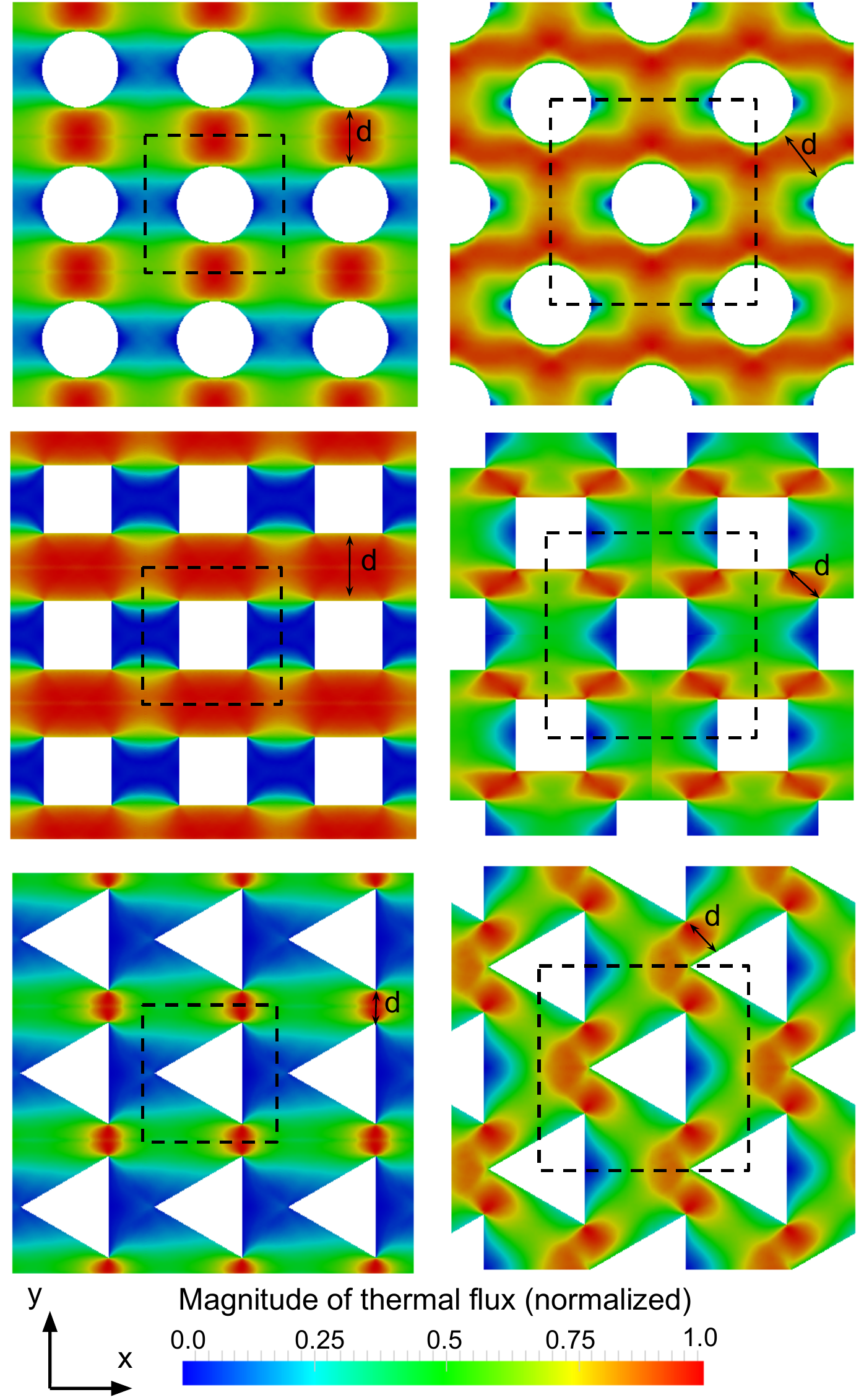}
\caption{Magnitude of the thermal flux for CA (upper left), SA (upper center), TA (upper right), CS (lower left), SS (lower center) and TS (lower right). The periodicity is $L = 10nm$. Red and blue areas refer to high and low thermal flux, respectively. Most of the heat flux travels away from the pores, due to phonon size effects. The unit cells as well as the pore-pore distances are underlined.}\label{fig:thermal}
\end{center}
\end{figure}
In Fig.~\ref{fig:thermal}, we show the magnitude of the thermal flux for all the configurations considered here, where the unit cell is highlighted. The imposed temperature gradient is along the $+x$ direction. The higher thermal fluxes are in the regions representing a continuous path from the hot to the cold side. Intuitively, staggered configurations should give lower PTCs with respect the aligned case with the same porosity, because of the lower view factor. However, as can be seen in Fig.~\ref{Fig:kappa} which shows our computed PTC for each of these configurations, for circular pores the aligned and staggered cases have very similar PTCs, in good agreement with experimental results~\cite{song2004thermal}. For the square pore cases, the staggered configuration exhibits a lower PTC than the aligned case. Interestingly, for triangular pores even though the staggered case has no direct path between the hot and the cold side, it has a higher PTC than the aligned case, which has the lowest PTC among all configurations. 

\begin{figure}[htbp]
\begin{center}
\includegraphics[width=7cm]{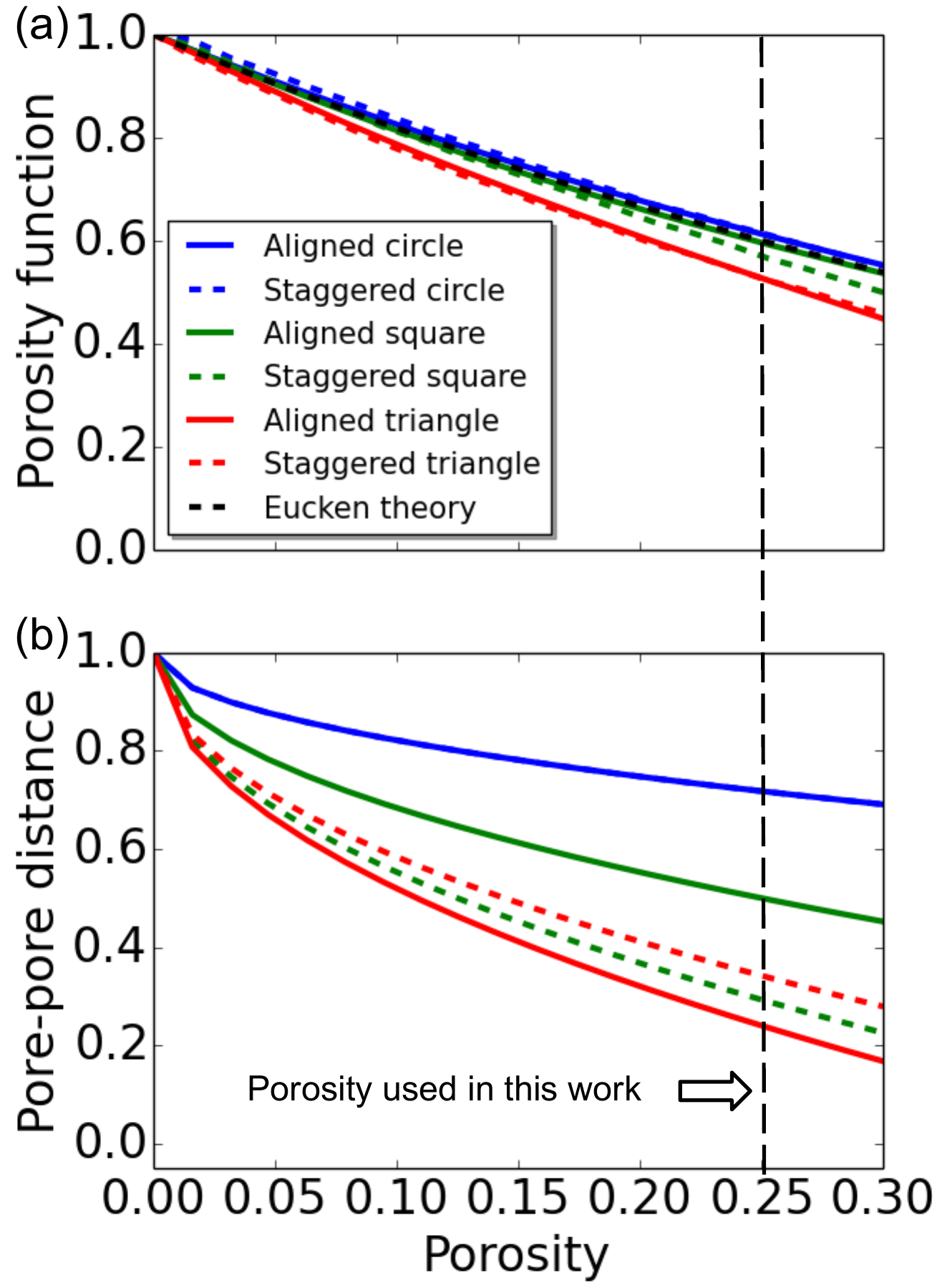}
\caption{ a) The porosity function versus porosity for different pore configurations, alongside the analytical model given by Eucken's theory. b) The pore-pore distance for different porosities. All the distances are normalized to the aligned unit cell size $L$.}\label{fig:porosity}
\end{center}
\end{figure}
To understand this trend, we first compute the effect of the porosity and the pore configurations at the macroscopic level. In absence of phonon size effects, heat transport is purely diffusive and the PTC can be described as $\kappa_{eff} = \kappa_{bulk} f(\phi)$, where $ f(\phi)$, the porosity function, is computed by Fourier's law. From Fig.~\ref{fig:porosity}-a, we note that the porosity function has relatively small variations for the six configurations at a given porosity, and can be fairly approximated by~\cite{hashin1962variational} $f(\phi) = \frac{1-\phi}{1+\phi}$. Furthermore, although configurations with triangular pores give lower porosity functions, there is basically no difference between the TA and TS cases, meaning that the porosity function alone cannot explain the trend in the PTCs at the nanoscale. 
Conversely, this trend can be explained examining the pore-pore distances, whose formulae are tabulated in~\ref{table:table1}.
\begin{center}
\begin{table}[htbp]
\begin{tabular}{ c || c lc l}
    & Circle & Square & Triangle \\ 
    \hline
  Distance (A) & $1-\sqrt{\frac{\phi}{\pi}}$ & $1-\sqrt{\phi}$& $1-\frac{2}{\sqrt[4]{3}}\sqrt{\phi}$\\
  Distance (S) & $1-\sqrt{\frac{\phi}{\pi}}$ & $1-\sqrt{2\phi}$ & $ 1- \sqrt[4]{3} \sqrt{\phi}  $\\
 \end{tabular}
 \caption{Pore-pore distance normalized to the size of the unit cell in the aligned configurations. For the TS case, only an approximate formula is reported. }
\label{table:table1}
\end{table}
\end{center}
In order to provide a common reference, all the pore-pore distances are normalized to the size of the unit cell $L$ for the aligned configurations. From Fig.~\ref{fig:porosity}-b we note that the trend in the pore-pore distance reflects well the trend in the PTCs. Indeed, for the TA case we have the lowest pore-pore distance. These results suggest that the pore-pore distance has a primary role in tuning PTCs, whereas the view factor plays a less important role, at least for the scale and geometry we have considered.   
\begin{figure}[htbp]
\begin{center}
\includegraphics[width=7cm]{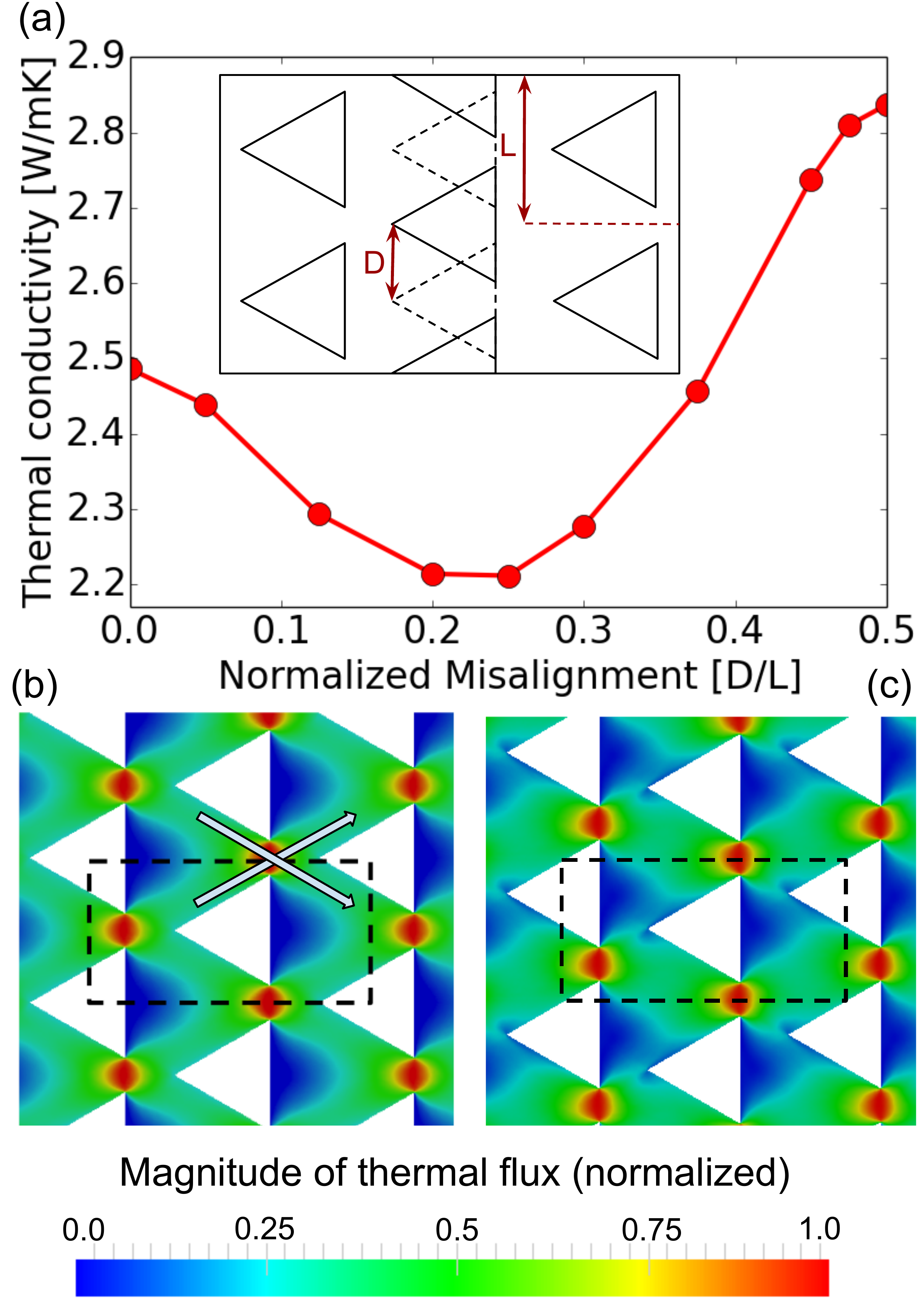}
\caption{(a). The PTCas a function of adjacent rows of triangular pores. The misalignment is normalized to the size of the unit cell for the aligned case $M=D/L$. Across each gap between two pores, two direct paths are available for phonons to travel ballistically between the hot and the cold side. (Lower panel) Magnitude of thermal flux for the $M = 0.5$ (b) and $M = 0.25$ (c) cases. }\label{Fig:kappa2}
\end{center}
\end{figure}

Motivated by these results we further explored the effects of pore-pore misalignment on the PTC for the triangular case by shifting the pore rows with respect to one another by distance $D$. The PTC as a function of misalignment M=D/L, is shown in Fig.~\ref{Fig:kappa2}-a, where $M=0$ corresponds to perfectly aligned pores and $M=0.5$ perfectly misaligned. The latter case refers to a staggered case with a rectangular unit cell, as shown in Fig.~\ref{Fig:kappa2}-b. By keeping the pore-pore distance fixed, we are able to isolate the effect of the view factor on the PTC. As can be seen in Fig.~\ref{Fig:kappa2}, for a normalized misalignment $M= 0.5$ each gap between pores contributes to two direct  and oblique paths from the hot and cold contacts, causing an increase in the view factor with respect to the TA case, where $M = 0$. For intermediate cases, the view factor vanishes and for $M=0.25$, reported in Fig.~\ref{Fig:kappa2}-c, we reach the minimum in the thermal conductivity, achieving a decrease in heat transport of more than $60\%$ with respect the CA case. We emphasize that, despite the fact that the two configurations have different PTCs, they still have the same porosity, and therefore we obtain an increase in ZT.


In summary, we have computed the PTC of np-Si with different pore arrangements and correlate the trend in PTCs with the geometric configurations. We deduced that the staggered configurations are not always the best choice for a given porosity and pore shape, and that a combined analysis based on both the view factor and pore-pore distance is needed. We identified the pore configuration that minimizes thermal transport, whose suppression ability is $60\%$ higher than that obtained by the simple circular aligned pore configuration. This result might serve as guidance for multiscale engineering of pore-boundary scattering in complex-shape materials which is crucial for the development of high performance thermoelectric materials.  

%

\end{document}